\documentclass{raa_twocolumn}            
\usepackage{graphicx,times}             
\usepackage{natbib}
\usepackage{amssymb,amsmath}
\bibpunct{(}{)}{;}{a}{}{,}
\usepackage[pagebackref=true]{hyperref}
\begin{document}

  \title{The SVOM / ECLAIRs Scientific Analysis Pipeline}

   \volnopage{Vol.0 (202x) No.0, 000--000}      
   \setcounter{page}{1}          

   \author{A. Goldwurm 
      \inst{1,2,*}\footnotetext{$*$ Corresponding Author: {\it andrea.goldwurm@cea.fr}},
    P. Bacon\inst{1}, N. Bellemont\inst{1}, F. Cangemi\inst{1}, C. Cavet\inst{1}, A. Coleiro\inst{1}, A. Foisseau\inst{1}, A. Gros\inst{3,**}, C. Lachaud\inst{1}, S. Le Stum\inst{1},
   T. Bouchet\inst{3}, J. Rodriguez\inst{3}, L. Bouchet\inst{4}
   }
   \institute{Universit\'e Paris Cit\'e, CNRS, CEA, Astroparticule et Cosmologie, F-75013 Paris, France; {$^*$\it 
   goldwurm@apc.in2p3.fr} \\
        \and
             CEA Paris-Saclay, Irfu, Département d'Astrophysique, F-91191 Gif-sur-Yvette, France;\\
        \and
             Universit\'e Paris-Saclay, Universit\'e Paris Cit\'e, CEA, CNRS, AIM, F-91191 Gif sur Yvette, France; {$^{**}$\it retired}\\
        \and
             IRAP, Université de Toulouse / CNRS / CNES, 9 Avenue du colonel Roche, F-31028 Toulouse, France.\\
\vs\no
   {\small Received 2025 December 15; accepted 2026 February 25}}

\abstract{This paper reports on the scientific pipeline for the analysis of the ECLAIRs data of the 
SVOM mission.
We describe the overall procedure, the different steps and the main algorithms 
of the data analysis for this hard X-ray coded mask instrument. 
The pipeline runs in automatic mode at the science center generating standard products 
but can also be used off-line with specific choices of parameters.
Generated data products and preliminary performances are illustrated 
along with perspectives for future improvements.
}
   \authorrunning{A. Goldwurm et al.}            
   \titlerunning{ECLAIRs Scientific Pipeline}    
\maketitle
%

\section{Introduction}\label{sect:intro}          

The ECLAIRs instrument \citep[][]{godet25} on the SVOM space mission \citep[][]{cordier25}
is an X-ray / hard X-ray (4-150 keV) telescope based on the coded-mask imaging technique \citep{golgro22}. 
A position-sensitive detector of 80~$\times$~80 square CdTe semiconductor pixels
records the shadows of a mask projected by the sources in the field of view (FoV).
A sky image is reconstructed by correlating the detector image to the mask pattern 
and by applying cleaning algorithms to reduce the systematic effects.
A dedicated "ECLAIRs pipeline" (ECPI) to perform such an analysis 
has been developed in the frame of the SVOM French Science Center (FSC) \citep{louvin25}.

As the set of X-band telemetry (TM) data, 
received on ground during a mission download period (pass),
are pre-processed at the FSC, ECPI runs automatically in quick look analysis (QLA) mode, 
feeding the SVOM science database (SDB) with the generated science data products (SDP)
to be distributed to the scientists. 
The pipeline can also be used off-line for a specific analysis
on selected data and with user-chosen parameters (energy bands, time intervals, etc.), 
allowing also the combination of the derived products (stacking, mosaics).
The generated SDP are stored in FITS files,
compatible with most of the standard astronomical tools (e.g., FTOOLS, DS9, XSPEC) and
including all needed metadata (energy bands, exposure times, etc.) in keywords (KW).

We describe here the main steps and algorithms of the analysis,
mentioning the acronyms of the pipeline components and of the SDP files.
The analysis principles are explained in \citet{golgro22}
and the treatment is similar to the one applied to INTEGRAL~/~IBIS data \citep{goldwurm03}
as the instruments are similar. 
Notable differences are that the ECLAIRs FoV is much larger ($\approx$ 90°$\times$~90°),
the Earth often obscures part of it, due to the mission observation strategy, 
the angular resolution ($\approx$~1.5° FWHM) and location accuracy ($<$~12$'$) are lower,
and the background is highly variable, due to the low orbit and 
frequent passages of the satellite in the energetic particle zones of the South Atlantic Anomaly (SAA).

   \begin{figure*}
   \centering
   \includegraphics[width= 1\textwidth]{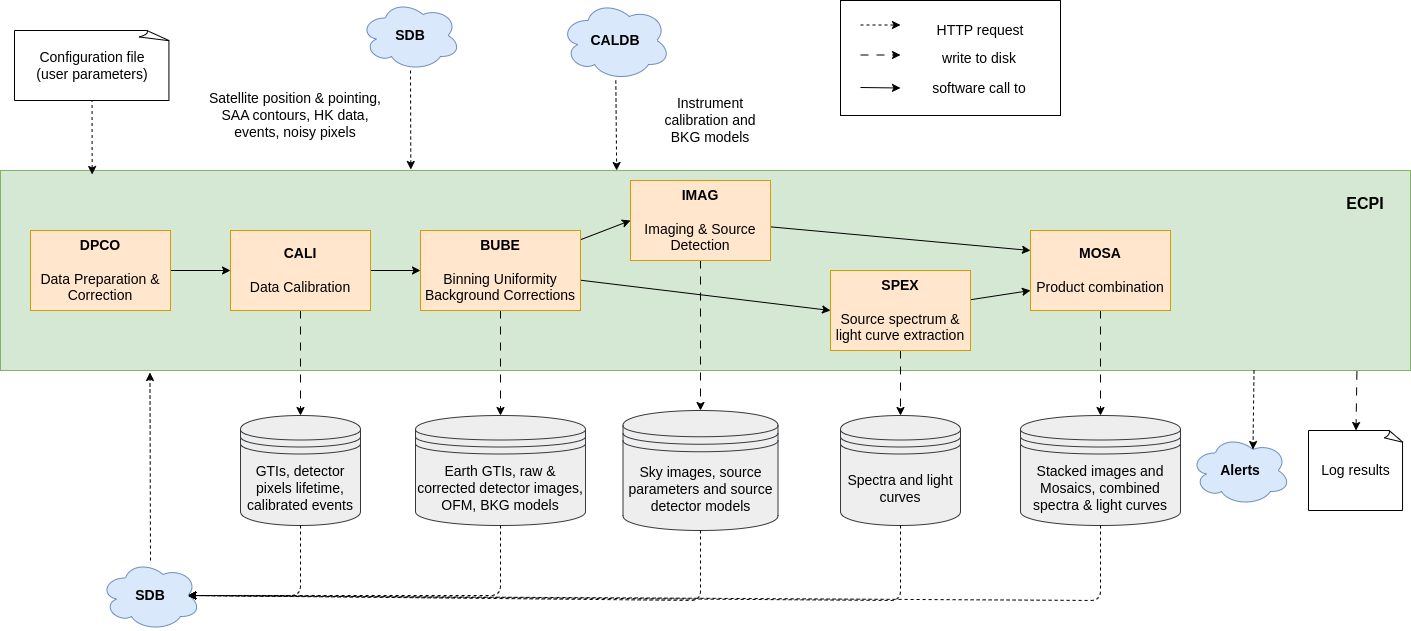}
   \caption{
   {\small Scheme of the ECLAIRs scientific analysis pipeline.
   Processing components are shown as rectangles in a sequence from left to right.
   Input preprocessed TM data are read from the SDB, the auxiliary data
   (calibration, response and model) from the CalDB.
   SDP generated by the different components (shown as cylinders in the bottom)
   are stored in the SDB and may be in input to the following components.
   The MOSA component is run on SDP produced by the previous components for different passes or observations.
   }
   }
   \label{Fig:Pipeline}
   \end{figure*}

\section{Overview of the pipeline} \label{sect:over}

The scheme of the pipeline is sketched in Fig.~\ref{Fig:Pipeline}.
It has as inputs the SVOM preprocessed TM data stored in the SDB 
and the set of correction, calibration and response files, 
prepared by the ECLAIRs Instrument Center
and stored in the calibration database (CalDB).
ECPI is composed of six large components
that run in sequence, with some loops described below.
Each component generates SDP, stored in the SDB, 
which are used by the following components and/or directly by the scientists.

The first two of them prepare the data and the correction parameters (DPCO) 
and calibrate and flag the events (CALI).
They can be run only once, and in the most general case, the user can start the analysis  
from the following component (BUBE), using the SDP stored in the SDB by the automatic processing, 
since no choice of scientific parameters is made before this level. 

In the BUBE component (Binning, non-Uniformity, Background, with Earth occultation, correction), 
based on the input user parameters,
the events are selected and binned in detector images, 
which then are corrected for non-uniformity and background modulation, 
considering the Earth obscuration.
Sky images are then reconstructed and sources are detected, positioned, identified with a catalog source
and their ghosts cleaned, in the so called IMAG (imaging) component.

To obtain products of specific sources, the BUBE component can  
be run again to bin detector images properly (small energy bins or time intervals), 
and then the source product extraction component (SPEX) generates source products, 
spectra or light curves (LC).
The different sets of SDP (for passes or observations) can be combined 
by stacking detector images and running the analysis components on them, 
by making sky image mosaics (component MOSA) and combining source spectra, LC and event lists. 

The ECPI pipeline is implemented in Python and is deployed as a Docker image within the FSC. 
In online mode, it interacts through an API interface with the other services (databases, pipelines, orchestration). 
On top of the scientific modules depicted in Fig.~\ref{Fig:Pipeline}, there are other software layers, 
that deal with data forwarding between scientific components, scientific data product generation and 
dynamic handling of the ECLAIRs calibration. These layers will be described elsewhere.

\section{Data correction and calibration} \label{sect:dpco}

The preprocessed ECLAIRs science data, in input to the data preparation and correction component (DPCO),
are in the form of event lists (EVT-CNV files), giving
for each recorded single event (multiple are neglected): the time (in UTC), the position Y-Z in the detection plane  
and the energy channel.
Other input TM data are the satellite attitude and orbit,
the pixels identified as noisy by the on-board software with the time they were set off/on,
and the instrument House-Keeping (HK) data (temperatures, tensions, etc.).
ECPI also uses the pre-calculated times of entry and exit of the SAA, 
both for the core and extended zones.

\subsection{Good Time Intervals and Dead/Noisy Pixels}
DPCO defines the good time intervals (GTI) to be used in the analysis,
and establishes the effective live-time of the pixels (PIX-LIF) during the period
of the input data. 
The GTIs are for periods out of SAA passages, 
of stable attitude (versus unstable or slewing intervals), 
of no or partial Earth occultation, during no TM gaps and good instrument functioning. 
Pixel life-time is computed from TM information on the noisy pixels and HK status, 
and using the CalDB tables of dead pixels (set off permanently).
The process also analyzes the science data in order to identify
residual dead or noisy pixels, not detected by the on-board software,
and to optimize the contours of the SAA based on the recorded count rate.
All extracted information is coded in the GTI and PIX-LIF files.

\subsection{Event Energy Calibration and Flagging}
The calibration component (CALI) performs the energy calibration of the events by applying 
a pixel gain and offset, tabulated in a dedicated calibration file,  
to the recorded channel (Pulse Height Amplitude, PHA) 
so that an invariant channel (Pulse Invariant, PI), 
with direct correspondence to an energy interval in keV, can be attributed to each event. 
The component also flags the events that are not valid 
either because they are out of the valid energy thresholds 
or because they are detected by pixels in periods of noisy status 
or with out-of-limit value of an HK parameter.
The calibrated and flagged events are stored (EVT-CAL) along with their GTI and PIX-LIF files. 

\section{Image Binning and Background and Non-Uniformity corrections} \label{sect:bube}
The first component of the proper scientific analysis (BUBE) bins the events 
in detector images for the requested time periods and energy channels
and correct them for the camera non-uniformity and non-flat background.

\subsection{Detector Image Binning and Efficiency Computation}
The procedure selects the "good" (flag = 0) events that  
belong to the user-requested GTIs, typically in periods out of SAA, 
with stable attitude, no TM gaps and when the FoV is not obscured by the Earth (NEO).
Partial (PEO) and total (TEO) Earth occultation GTIs can also be used, 
as well as other specific intervals, 
like the active periods of a transient source or even slew periods,
which must be entered by the user.
Events are binned into detector images generating, for each requested energy band, 
three 80$\times$80-element arrays (DET-IMA):
one containing the counts per pixel, one the associated variance 
and one the efficiency. 
The normalized efficiency of each pixel is inferred
from the PIX-LIF information and the ground-calibrated
pixel low and high energy thresholds. 
The user can also neglect, in the analysis, 
other parts of the detector that show malfunctioning 
(noisy pixels/sectors, column effect, etc.) 
by setting the corresponding pixel efficiency to 0.

\begin{figure}[ht]
  \centering
    \includegraphics[width=27mm,angle=0]{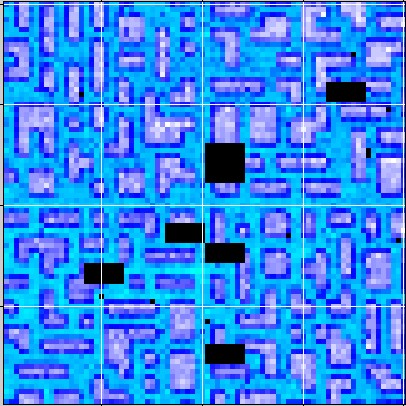}
    \includegraphics[width=26.5mm,angle=0]{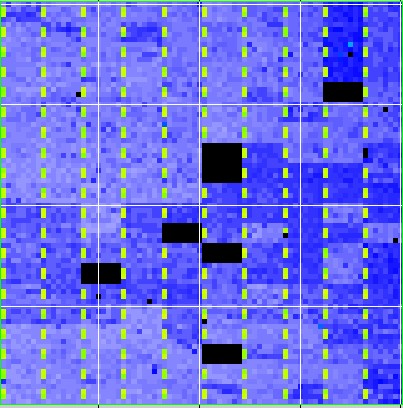}
    \includegraphics[width=27mm,angle=0]{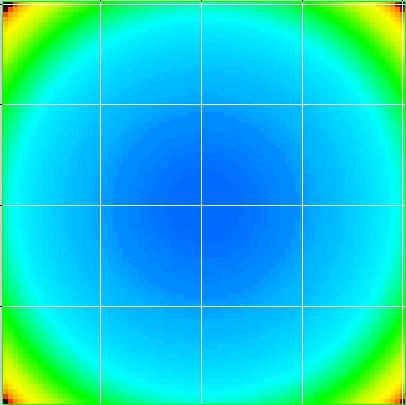}
    \caption{\label{Fig:Det-Eff-CXB}{\small 
    Left: A corrected intensity detector image during a Crab on-axis observation. 
    Center: The corresponding global efficiency image.
    Right: The background model detector image, 
    dominated by the CXB, not modified by the Earth occultation (NEO period).}}
\end{figure}

\subsection{Non-Uniformity and Background Correction with Earth Modulation}
While a coded mask system is not sensitive to temporal variations of the background or of the detector efficiency, 
any spatial modulation that is not produced by the mask hampers the proper reconstruction of the
source fluxes and positions \citep{golgro22}.
It is therefore necessary to correct the non-uniformity of the detector plane response \citep{xie24}
and the spatially modulated background, before decoding the sky with the mask pattern.
The correction of the detector image $D$ affected by a non-uniformity $U$ and a non-flat background $B$ 
can be performed by computing the corrected image as 
$D^C= \frac{D}{E \cdot U} - b \cdot B $
where dot or division means element-by-element matrix multiplication or division,
$E$ is the efficiency and the scalar $b$ a normalization factor, 
presently computed from the exposure time, but in the future possibly fitted to the data.
The variance is computed accordingly. 
The efficiency image associated to $D^C$ is now the global efficiency $E^g=E \cdot U$. 
The main contribution to the background is the isotropic Cosmic X-ray Background (CXB)
and its model is calculated from the solid angle that each pixel of the camera subtends to the sky
considering the ECLAIRs geometry and the CXB spectrum \citep[][]{moretti09}.
When the FoV is obscured by the Earth, the induced modulation on  
the CXB shape during the PEO GTI is taken into account.
The corrected detector images (DET-UBC) 
and the background detector models used for the correction (DET-BKG) (Fig.~\ref{Fig:Det-Eff-CXB})
are saved in outpput, along with the computed time-averaged FoV non-obscured by the Earth (EAR-OFM)
(Fig.~\ref{Fig:Products} bottom-right).

\section{Image Reconstruction and Source Localization} \label{sect:imag}
\begin{figure}[ht]
       \centering
          \includegraphics[width= 1.\linewidth]{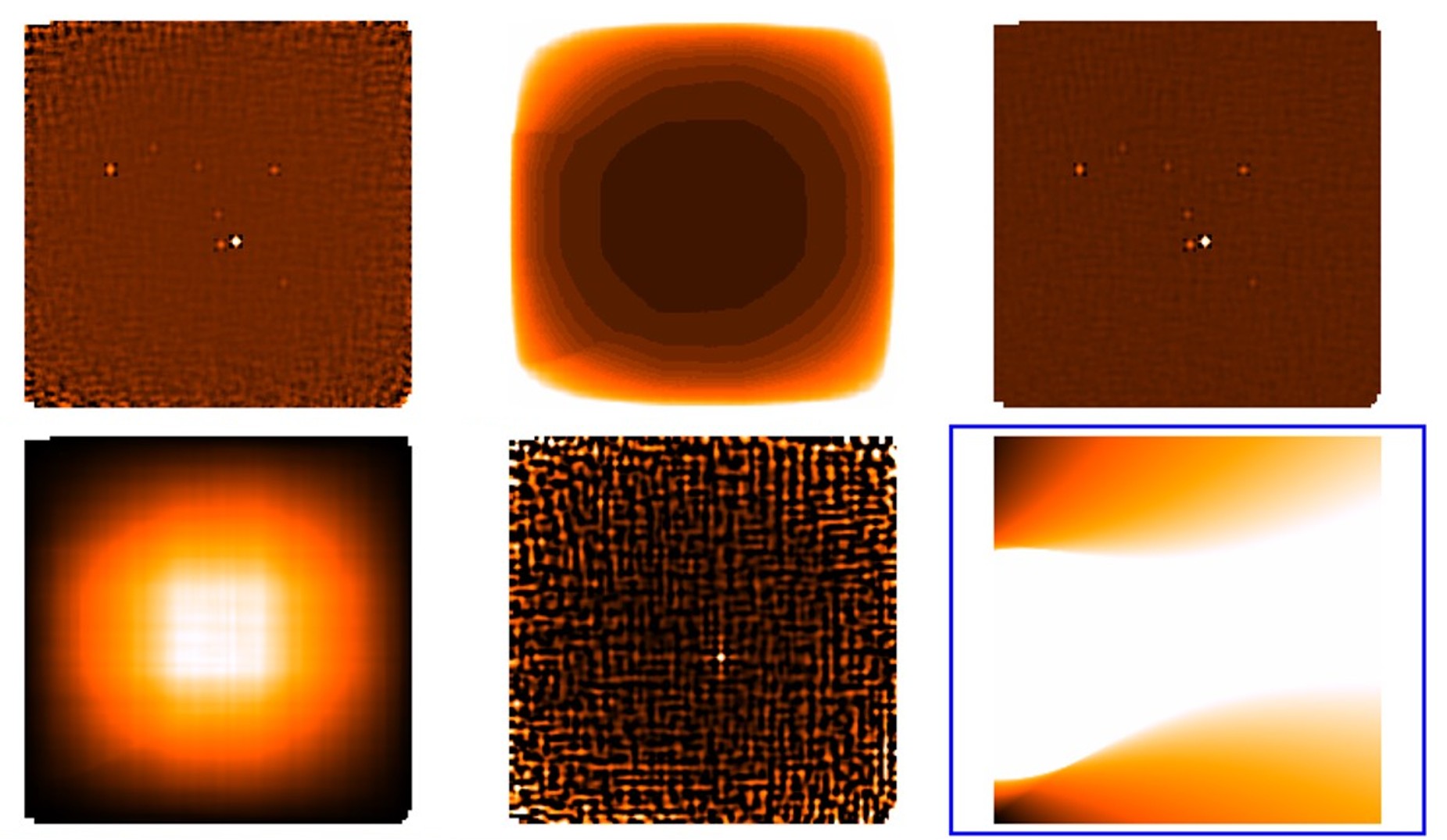}
           \caption{\label{Fig:Products}{\small 
           IMAG sky image reconstruction products. 
           From top-left to top-right: Intensity, variance and SNR images;
           bottom-left and bottom-center: effective exposure time and sum of deconvolved source models.
           The bottom-right image is the Earth open FoV map computed by BUBE.} }
    \end{figure}

\begin{figure}[ht]
       \centering
     \includegraphics[width=0.98\linewidth]{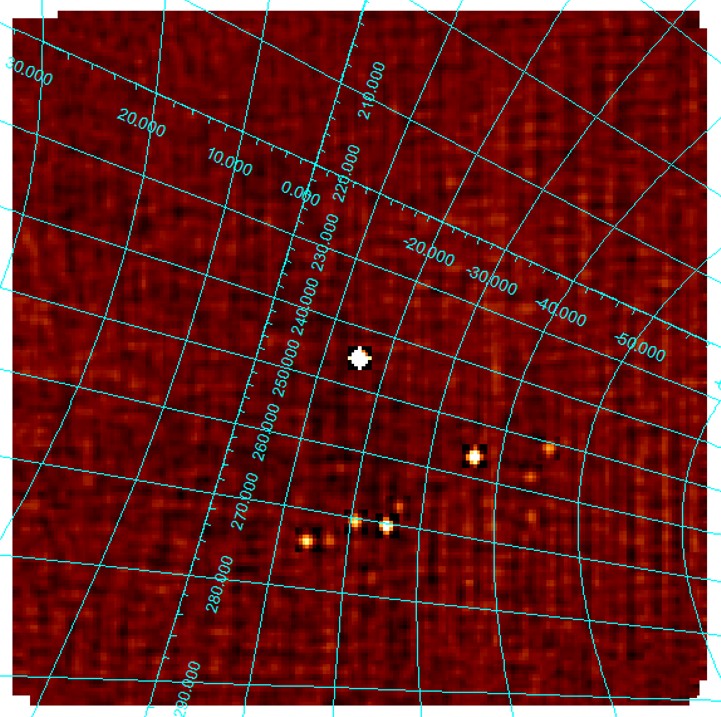}
          \caption{\label{Fig:Sky}{\small Reconstructed SNR sky image of the Sco~X-1 field,
          shown in logarithmic scale with equatorial coordinates: 
          several sources of the galactic plane are visible below the bright central target.} }
  \end{figure}

Reconstruction of sky images (IMAG component) 
is performed by deconvolving the corrected detector images, 
iteratively searching and localizing sources, modeling and subtracting their secondary lobes,
then correcting for Earth occultation and off-axis effects, before sky map normalization 
and unit transformation (Fig. ~\ref{Fig:Products} - ~\ref{Fig:Sky}).

\subsection{Sky Image Deconvolution}
A sky image is derived from the detector image $D$, its variance $D_V$ and efficiency $E$, 
by decoding it with the mask pattern following the prescriptions of \citet{golgro22}. 
The mask of $46\times46$ quasi-random transparent and opaque elements plus a central opaque cross \citep{lachaud25}
(represented by 0. value when opaque and 1.0 when transparent) 
is re-binned over an array of pixels of same area as the detector pixels 
(linear ratio mask-element/detector-pixel = 2.53).
From the re-binned mask array $M$ of $120\times120$ pixels with values between 0.0 and 1.0.
we define the decoding array 
$G =  \frac{1}{a} \cdot M - \textbf{1}$
where \textbf{1} is an arrays of 1s and the scalar $a$ is the aperture of the mask (=~0.4).
Using 
$
G^+ = \begin{cases}
    G \rm{~~for~} G \ge 0 \\ 
    0 \rm{~~elsewhere} \\
\end{cases}
$ and
$
G^- = \begin{cases}
    G \rm{~~for~} G \le 0 \\ 
    0 \rm{~~elsewhere} \\
\end{cases} 
$
the sky image $S$ is given by
$$
    S = \frac{G^+ \star (D \cdot W) - Bal\cdot (G^- \star (D\cdot W))}{A}
$$
where $\star$ means correlation operator. 
The weighing array can be taken as $W = E$, 
neglecting in this way the pixels that are off and giving low weight to those with low efficiency,
but can also integrate other corrections, 
for example to neglect detector parts contaminated by bright sources or high background.
The balance array 
$ Bal = \frac{G^+ \star W }{G^- \star W } $
gives the open to closed mask element ratios and ensures a flat image with a mean of 0 in absence of sources. 
The normalization array
$ A = (G^+ \cdot M) \star W - Bal \cdot ((G^- \cdot M) \star W ) $
takes into account the partial modulation.
With this normalization the sky image $S$ of dimensions $199\times199$ pixels give 
at the source peak the mean recorded source counts within one totally illuminated detector pixel.
The associated sky variance $V$ is (where $X^{2} = X \cdot X$)
$$
V=\frac{(G^{+})^{2} \star (D_V \cdot W^2) + Bal^2 \cdot ( (G^{-})^{2} \star (D_V \cdot W^2) )}{A^2}
$$

Finally the signal-to-noise image is $SNR = S/\sqrt{V}$.


\subsection{Source Localization, Identification, Modeling and Iterative Cleaning, Sky Image Normalization}
The decoded sky image is searched for point sources, 
i.e. SNR excesses larger than a user-defined threshold of $\approx$ 6-10.
The source peak, of size (FWHM) equal to about the projected mask element, 
is fitted with a 2D-Gaussian, approximation of the spatial response, 
the so-called System Point Spread Function (SPSF) \citep[][]{lachaud25},
in order to estimate the source fine position.
The image coordinates are transformed in celestial ones and
the source is identified using a catalog of known X-ray objects.
Estimated source parameters (fine positions, fluxes, SNR, identified objects, etc.)
are provided in output (SOP-IMA).

Due to the quasi-random ECLAIRs mask pattern \citep{lachaud25}, the main source peak 
is surrounded by a "coding noise" composed of numerous smaller secondary positive and negative lobes. 
This noise is distributed all over the FoV, depends on the source location and intensity 
(see bottom-center of Fig. ~\ref{Fig:Products}), and can hamper the search of weaker objects. 
It is therefore necessary to implement an iterative cleaning procedure 
in order to subtract the secondary lobes of a source before searching for a weaker one.
Each detected source is modeled at the fitted position, 
the detector image model (DET-MOD) is decoded, normalized and
subtracted from the initial sky image, reducing the coding noise contamination 
before searching and deriving parameters for a weaker source.

At the end of the loop, source main peaks are reintroduced in the cleaned image, 
which is then normalized for the effect of the Earth occultation and off-axis effects 
and converted to units that give at each point the source count rate detected on the whole camera,
as for an on-axis location. 
The proper word coordinate system KWs (J2000 equinox equatorial coordinates in tangential projection),
computed from the GTI-average attitude, are associated to the image FITS file of the
cleaned and normalized sky images (SKY-IMA). 
Examples of in-flight data IMAG products are shown in Fig.~\ref{Fig:Products} 
and a cleaned SNR sky map of the Sco X-1 field in Fig.~\ref{Fig:Sky}. 

\begin{figure}[ht]
  \centering
    \includegraphics[width=1.0\linewidth]{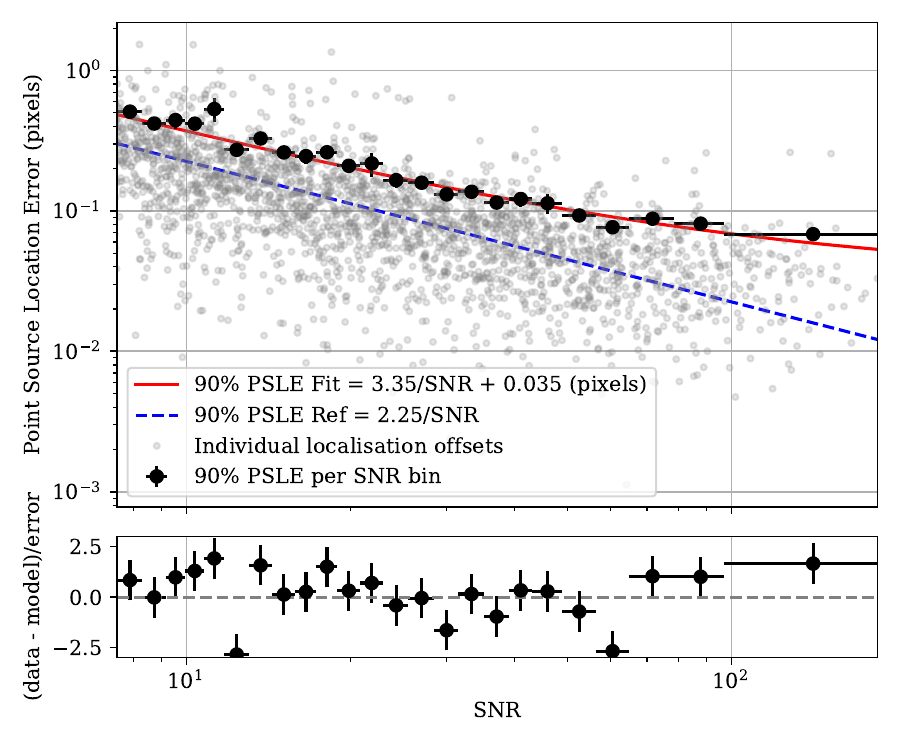}
	  \caption{\label{Fig:PSLE}{\small 
      Estimated ECLAIRs PSLE from 1 year of in-flight data of known-sources: measured localization offsets in pixel unit vs source SNRs (grey points), their 90$^{th}$ percentiles in the SNR bins that contain 100 points each (black points), the best fit function of the PSLE at 90\% c.l. vs the SNR (red line), compared to the the theoretical PSLE for a perfect system (blue broken line). Fit residuals are shown at the bottom.}
      }
\end{figure}

\subsection{Preliminary Estimation of Imaging Biases and Point Source Location Error}
Using results of the imaging analysis on ECLAIRs data of the first year observations
of known bright 
point-like sources, namely the offsets between the SPSF
fitted centroids and the known source positions,
we have studied the Point Source Location Error (PSLE) of the system, 
determined some of the imaging biases 
and then the parameters of the function PSLE versus (vs) the SNR, which characterizes
the location accuracy for a source of given SNR \citep[see][]{gros03, golgro22}.
We have derived a more precise "effective distance" between the mask and the detector plane,
which results in +0.11~cm with respect to the ECLAIRs geometry,
and a marginally significant misalignment with respect to the platform attitude
along the image axes (Y, Z) and in rotation of $\approx$ 0.7$'$, 0.9$'$, and 0.3$'$, respectively.

The preliminary calibrated PSLE, at 90$\%$ confidence level (c.l.), vs the SNR 
is illustrated in Fig.~\ref{Fig:PSLE},
once the estimated biases are corrected,
and it is compared to the expected trend in case of a perfect system with the same geometry as ECLAIRs. 
The PSLE was estimated by measuring the offsets 
of Crab, Cyg X-1, Cyg X-2, and Cyg X-3. After a selection of images containing no artifacts, 
the resulting 2475 data points were binned by SNR with 100 points per bin.
Then, to obtain a robust 90$^{th}$ percentile and the associated error bar in each bin, 
a bootstrapping procedure was applied: in each bin, 100 points were drawn with replacement a large number of times \citep[][]{foisseau25}. 
The mean 90$^{th}$ percentile and its standard deviation in all samples are used as data points and 
errors in the fit of the 90\% c.l. PSLE function. 
We find (Fig.~\ref{Fig:PSLE}) that the trend of the error, in units of sky  pixels, vs the SNR is given by 
$$
\textrm{PSLE}_{90\%}(\textrm{SNR}) = \frac{(3.35 \pm 0.03)}{\textrm{SNR}} + (3.5 \pm 0.5) \times 10^{-2} ~~pix
$$
Since, on-axis, a sky pixel corresponds to $\approx$34$'$, 
the error at the image center varies between 12.5$'$ to 2.3$'$ for SNRs between 10 and 100.
Note that due to the tangential projection the off-axis pixel angular sizes reduce along the direction 
from the center, to reach, at the edges of the FoV, about half the size the on-axis value. 
The fact that a plateau (1.2$'$ on-axis) is reached at high SNR indicates 
that some systematic effect is still present, while the error at low SNR may be dominated 
by noise in the images, as indicated by the out-layer points. 
Future improvements of the pipeline for the reduction of noise 
and the correction of residual imaging biases may reduce this localization uncertainty.

\begin{figure}[h]
  \centering
    \includegraphics[width=1.0\linewidth,angle=0]{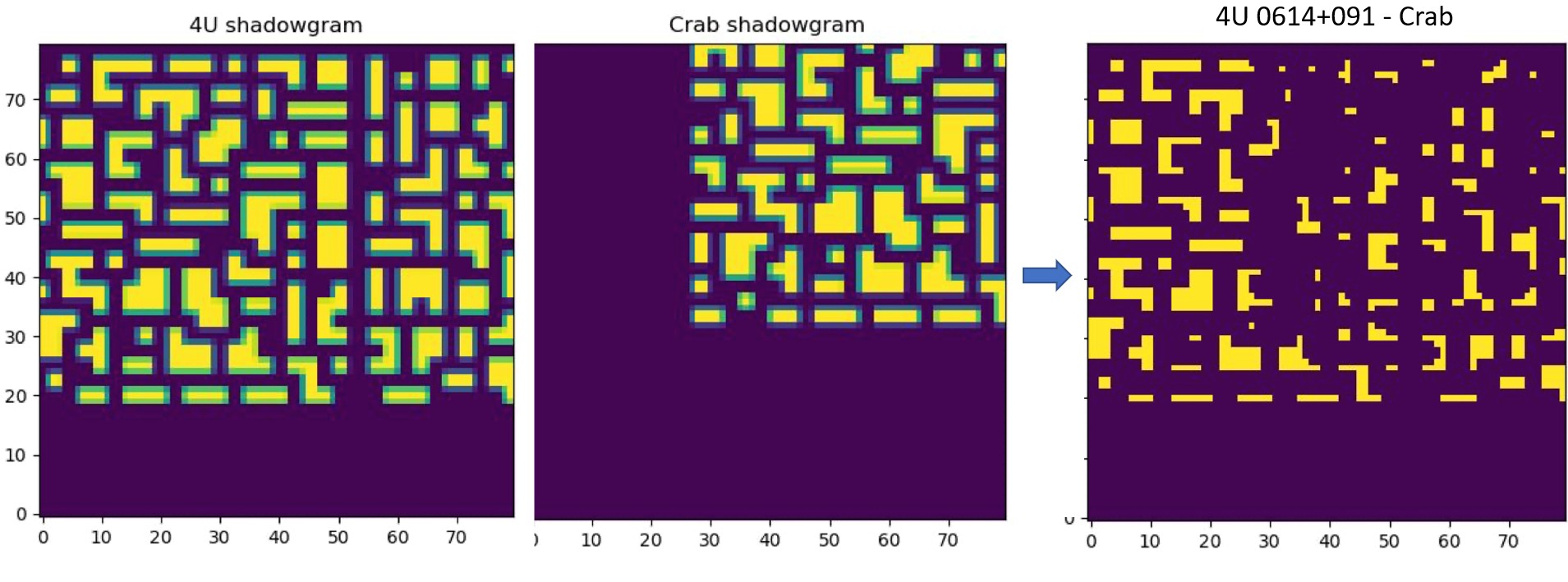}
    \includegraphics[width=1.0\linewidth]{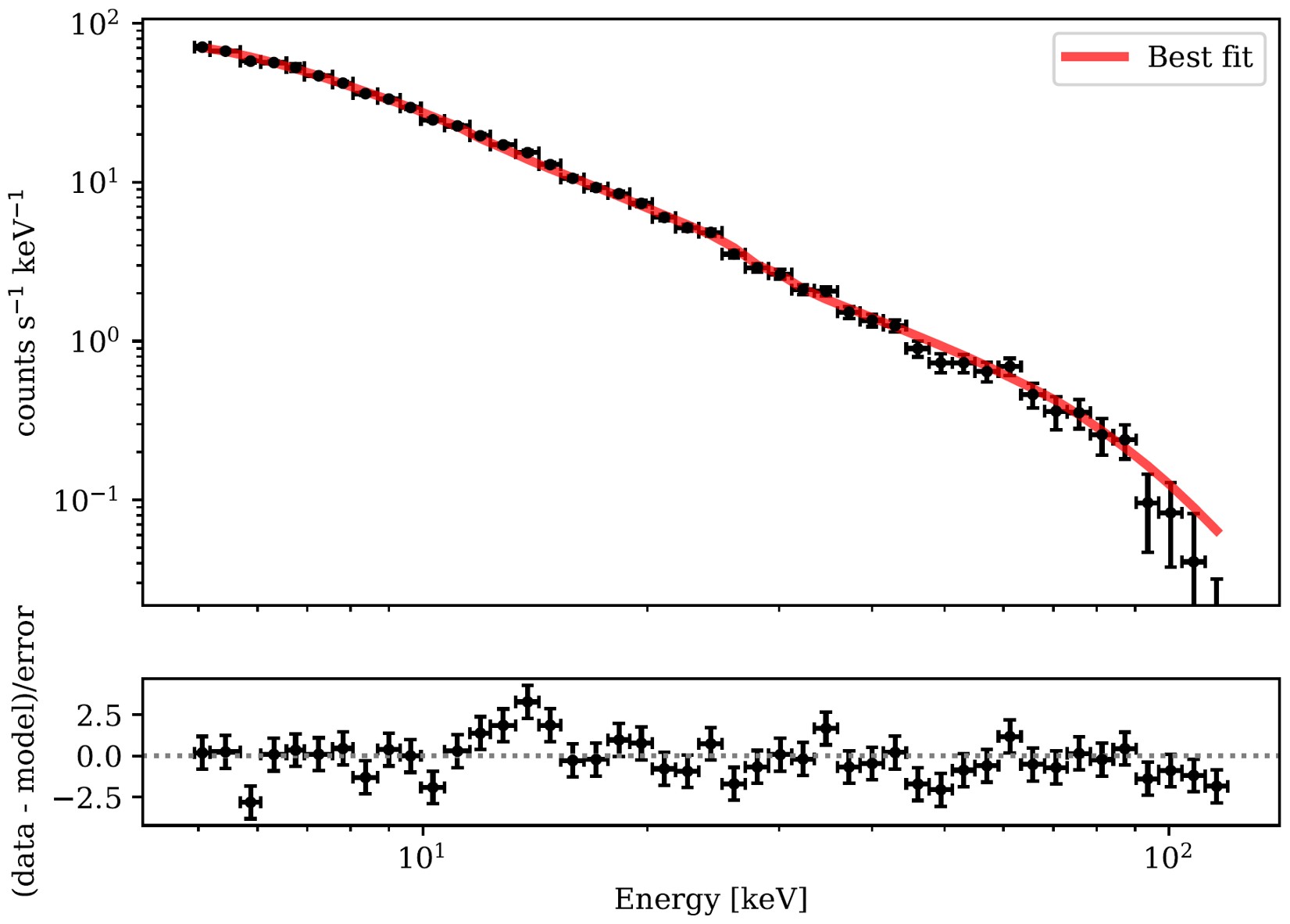}
	  \caption{\label{Fig:Crab-CtsSpect}{\small Top: Source detector image models for two sources and combination for event selection of the 1$^{st}$ one. Bottom: ECLAIRs count spectrum of the Crab with its best fit power-law model (red line) and residuals, for an on-axis NEO observation of $\approx$ 800 s.}}
\end{figure}
%
   \begin{figure*}
   \centering
   \includegraphics[width=1\textwidth, angle=0]{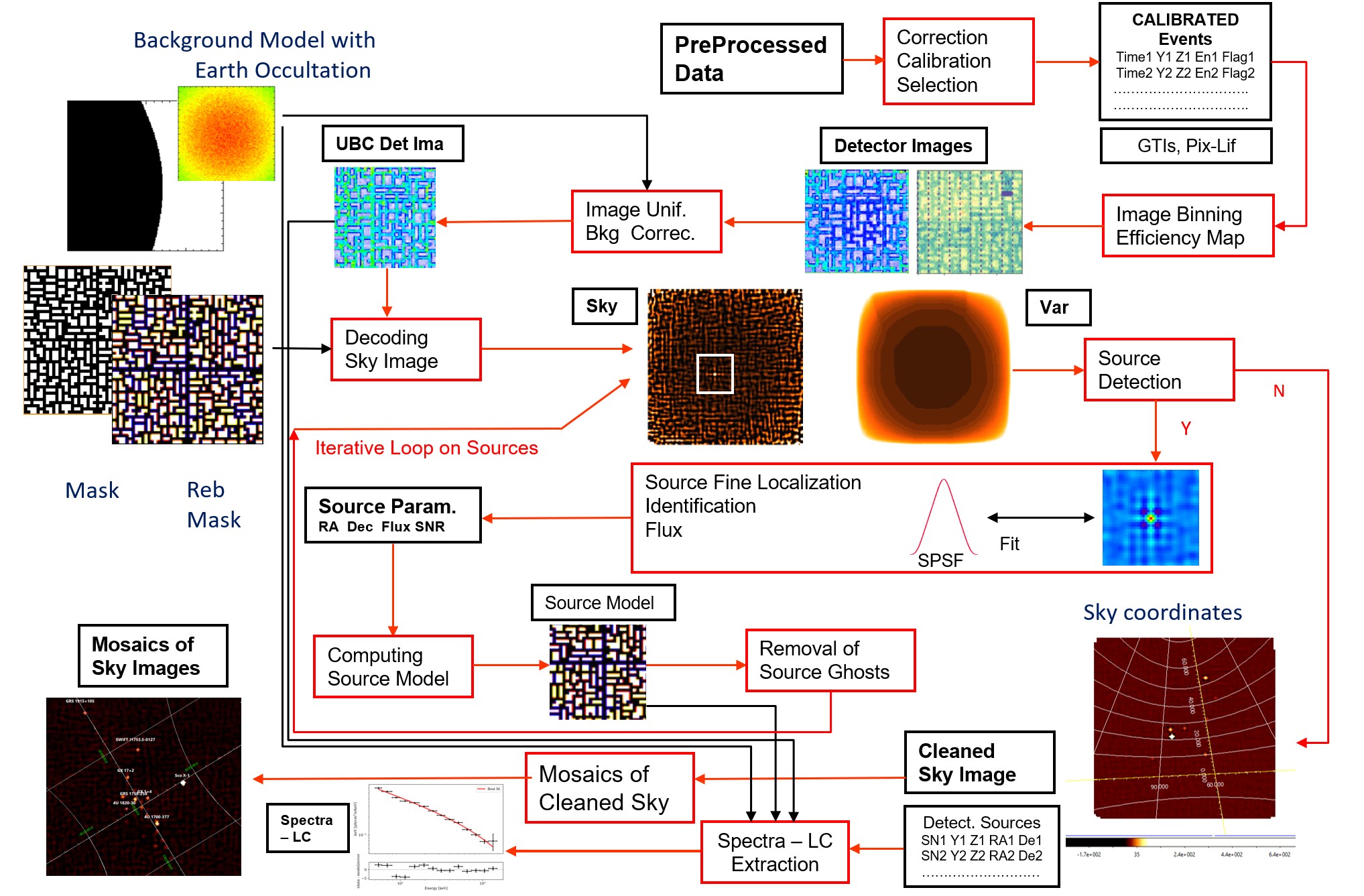}
   \caption{\small An illustrative scheme of the ECLAIRs data scientific analysis 
   logic, implemented by ECPI and associated software, using simulated data and derived products
   (adapted from \citet{golgro22}).}
   \label{Fig:Analysis}
   \end{figure*}

\section{Source spectra and light curves Extraction} \label{sect:spex}

Once the positions of the active sources in the FoV are defined,
either by the imaging process or by a choice of the user, 
the extraction of source products (SPEX) can be performed. 

Data are reprocessed with BUBE to bin and correct detector images in the desired energy or time bins
(in automatic QLA mode at FSC the same binning of 5 energy band of imaging is used 
and SPEX in spectral mode is called directly after IMAG without running again BUBE).
They are used to fit a global model that contains the computed shadow-grams 
(Fig.~\ref{Fig:Crab-CtsSpect} top)
of the chosen sources (at their fixed positions) and, possibly, a background model. 
The fit can be performed by least-squares (default) 
or maximum likelihood (adapted to very short exposures) methods,
and determines simultaneously the fluxes of all sources and of the background.
The normalized to on-axis count rate spectra (SPS-SOE) or LC (LCS-SOE)
contain fluxes and errors for all fitted sources and, if included, the background.
The spectra (Fig.~\ref{Fig:Crab-CtsSpect} bottom), 
after some manipulations, can be compared, using standard tools like XSPEC,
to spectral models convolved with the instrument energy response (ENE-RSP), 
described by an ARF and RMF matrices and the PI-channel to energy relation.
Note that the background spectrum is already subtracted by the procedure from 
the source spectra, it shall not be subtracted again with XSPEC or other tools. 

At present the source model is simply the illumination of the detector 
given by the projection of the mask on the detector grid.
A more refined calculation considering instrument geometry and material transparencies 
is in preparation. 
However, especially at low energy (4-60 keV), the present approximation 
is sufficient, as the dominant interaction is photoelectric absorption 
and the mask has been designed to minimize the vignetting effect \citep{lachaud25}.

The user has the choice to fit the DET-UBC images, already corrected for background,
in which case a background spectrum is not generated in the output, 
or to fit as well a background model to uncorrected images (DET-IMA).
Such a model is, by default, a 2D polynomial function, whose parameters
are adjusted to the image simultaneously with the amplitude of the sources.
Other models (e.g., from empty field data or including Earth albedo and reflection) 
will be possibly implemented in the future.

\section{Spectra and LC Combination, event selection, stacking and mosaics} \label{sect:mosa}

The final component (MOSA) (Fig.~\ref{Fig:Pipeline}) and other tools, 
some provided as a separate software set from ECPI \citep{coleiro25},
are, or will be soon, available, to combine, handle and analyze the SDP.

In particular spectra (or LC) of a source derived from different observation periods
can be combined to obtain an average (concatenated) spectrum (LC),
improving the SNR (time coverage).
Spectra are provided in format compatible with XSPEC, for spectral modeling analysis.
These tools are also used to treat GRB slew data, when the attitude is not stable, as
the satellite is moving to point the triggering source. 
In this case the detector image binning and
spectral extraction are performed in several short sequences ($\approx$ 2~s) during which 
the attitude is approximately constant and then the derived spectra can be combined.
Another program performs the selection of calibrated events based on a source model image (DET-MOD),
optimizing the SNR of the event list for a given source
and reducing contamination of other sources of the FoV (Fig.~\ref{Fig:Crab-CtsSpect} top). 
After conversion of event times to the solar system barycenter
these source-optimized lists can be used for fine temporal studies \citep{lestum26}. 
Stacking of detector images, before decoding and analysis, for periods of constant attitude,
is already operational for NEO and PEO data, producing routinely the combined NPO products,
while stacking date of different passes of the same observation will be available soon
and will generate stacked products (STK files). 
Production of mosaics of sky images of different attitudes will also be soon implemented.

A simplified, logical scheme of the ECLAIRs data scientific analysis, that summarizes the described procedures, 
including the iterative loop to search and fit the sources and clean their coding noise, is illustrated 
in Fig.~\ref{Fig:Analysis}. 

\begin{figure}[ht]
  \centering
    \includegraphics[width= 1.0 \linewidth]{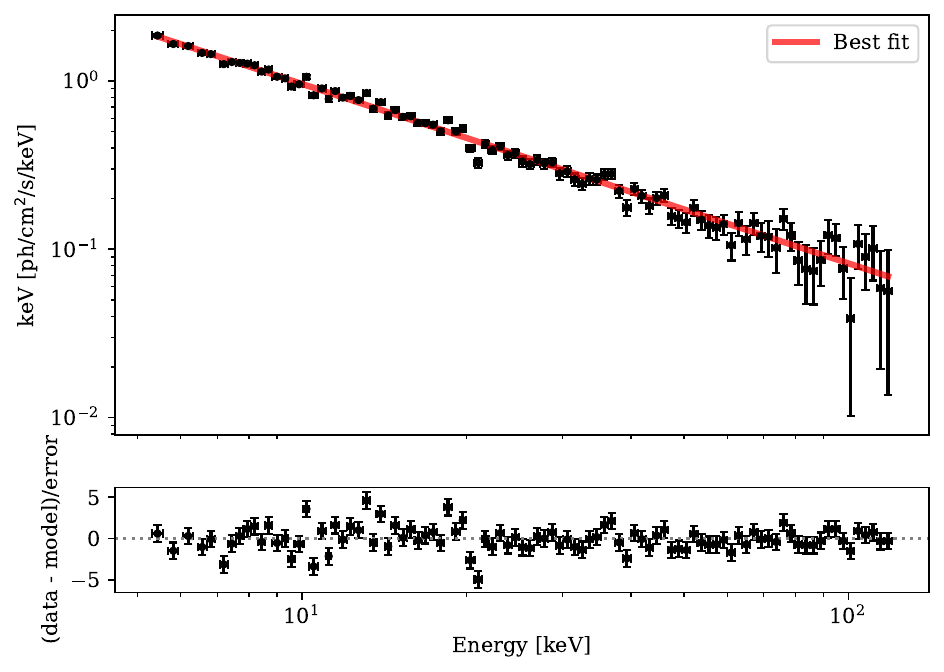}
	  \caption{\label{Fig:Crab-Spe}{\small 
      ECLAIRs photon spectrum of the Crab Nebula and its best fit power-law model (red line)
      from a $\approx$~970~s observation with the target off-axis ($\approx$~30°), with residuals in the bottom.}}
\end{figure}

\section{Pipeline performance and perspectives} \label{sect:conc}

ECPI runs automatically at FSC since the SVOM launch (June 2024) and feeds the SDB with SDP 
that are available to scientists, with standard choices of parameters 
(e.g., five energy bands, standard GTIs). 
Revised set of SDP will be regularly produced by the FSC with improved software and 
auxiliary files.
The ECPI is documented and can be installed in individual computers to run specific analyses.

\begin{figure}[ht]
  \centering
   \includegraphics[width= 0.95\linewidth]{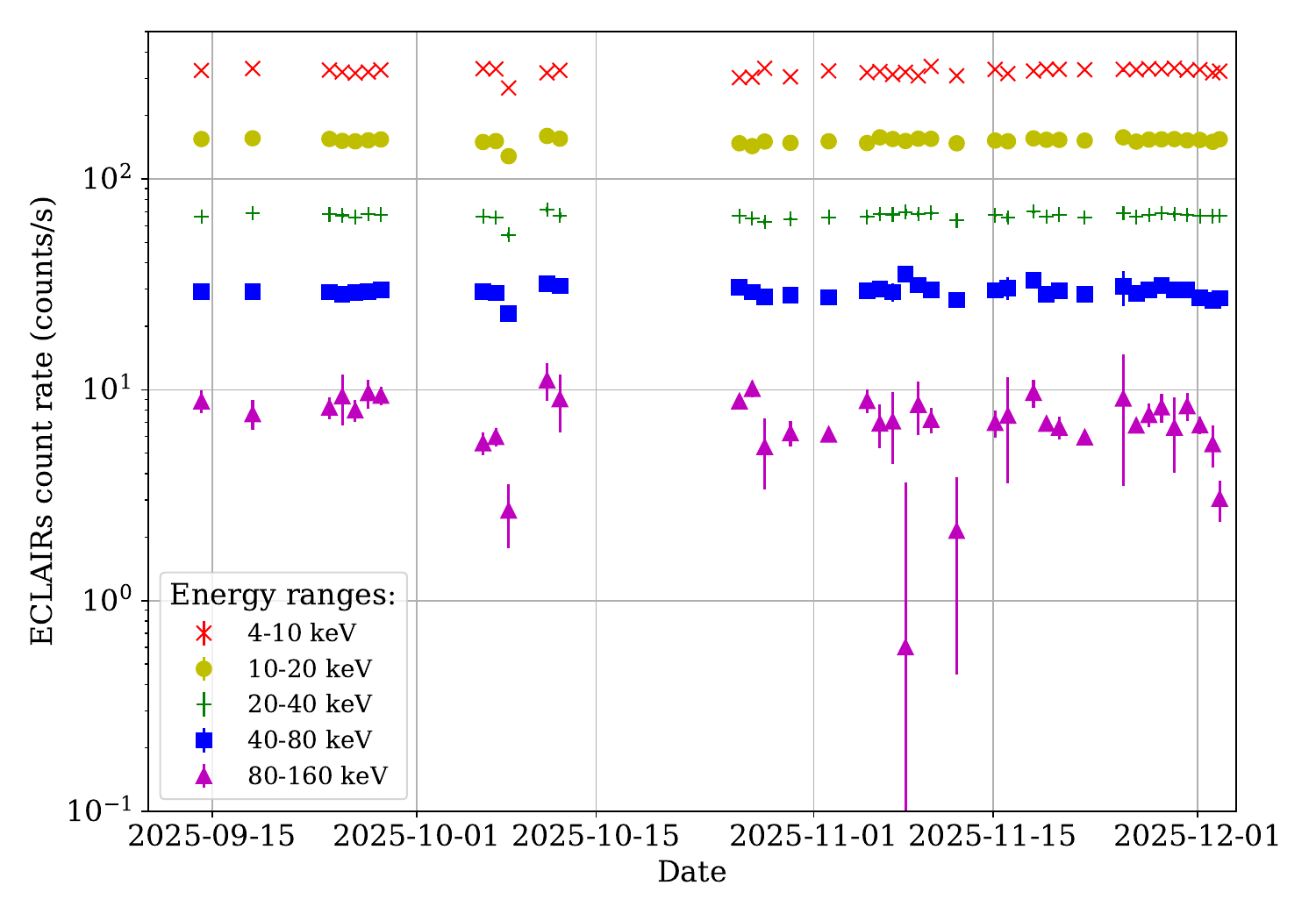}
	  \caption{\label{Fig:Crab-LC}{\small 
      ECLAIRs count rate light curves of the Crab Nebula in the fall of 2025 in different energy bands.}}
\end{figure}

The preliminary evaluations of the ECPI performance show that, overall, it works well. 
A set of scientific results have been obtained with it,
for both core and observatory science programs \citep[see][]{coleiro25},
and for the performance of ECLAIRs \citep{godet25}.
Even if more work is needed to better clean images from the background
(the correction with a 2D polynomial function fitted to the data, as for spectral extraction, is in progress), 
the source localization approaches the expected theoretical performance and is within the requirements,
provided that the images are not too affected by noise. 
The reconstructed spectra (Fig.~\ref{Fig:Crab-CtsSpect} and \ref{Fig:Crab-Spe}) 
and light curves (Fig.~\ref{Fig:Crab-LC}) of the Crab nebula,
the standard candle in X-ray astronomy, show that expected spectral parameters
and stable behavior are obtained when the background is well corrected, 
even for highly off-axis targets. 
ECLAIRs and MXT (the SVOM X-ray telescope) spectra combine very well \citep[][]{coleiro25} and 
comparison with results from other observatories also shows consistent trends.
The ECLAIRs scientific performance, 
from in-flight data analyzed with ECPI and other software, is presented in \citet{godet25}.

The work to complete ECPI functionalities (stacking, short-bin LC, mosaics, slew data),
to improve instrument and background modeling 
and complete the software tools to handle and analyze the products, is in progress.

\begin{acknowledgements}
The Space-based multi-band astronomical Variable Objects Monitor (SVOM) 
is a joint Chinese-French mission led by the Chinese National Space Administration (CNSA), 
the French Space Agency (CNES), and the Chinese Academy of Sciences (CAS). 
We gratefully acknowledge the unwavering support of NSSC, IAMCAS, XIOPM, NAOC, IHEP, CNES, CEA, and CNRS.
We thank A. Baudiquez, C. Catalano, J.-M. Colley, F. Dodu, H. Jimenez-Perez, D. Thibaut, who contributed to ECPI in the past.
\end{acknowledgements}





\begin{thebibliography}{99}

\bibitem[Coleiro et al. (2026)]{coleiro25} Coleiro, A., et al., 2026, RAA, this issue, xxx

\bibitem[Cordier et al. (2026)]{cordier25} Cordier, B., et al., 2026, RAA, this issue, xxx

\bibitem[Foisseau (2025)]{foisseau25} Foisseau, A., 2025, PhD Thesis, Universit\'e Paris Cit\'e, (\href{https://theses.fr/s347692}{theses.fr/s347692})

\bibitem[Godet et al. (2026)]{godet25} Godet, O., et al., 2026, RAA, this issue, xxx

\bibitem[Goldwurm et al. (2003)]{goldwurm03} Goldwurm, A., David, P., Foschini, L., et al., 2003, \aap, 411, 223

\bibitem[Goldwurm \& Gros (2022)]{golgro22} Goldwurm, A., Gros, A., 2022, Coded Mask Instruments for Gamma-Ray Astronomy in Handbook of X-ray and Gamma-ray Astrophysics. Eds. Bambi C., Santangelo A., Springer, Singapore, id.15, doi: \href{https://doi.org/10.1007/978-981-16-4544-0_44-1}{10.1007/978-981-16-4544-0\_44-1}

\bibitem[Gros et al. (2003)]{gros03} Gros, A., Goldwurm, A., Cadolle-Bell, M., et al., 2003, \aap, 411, 179

\bibitem[Lachaud et al. (2026)]{lachaud25} Lachaud, C., et al., 2026, RAA, this issue, xxx

\bibitem[Le Stum et al. (2026)]{lestum26} Le Stum, S., Cangemi, F., Coleiro, A., et al., 2026, \apj ~Lett, 997, L25

\bibitem[Louvin et al. (2026)]{louvin25} Louvin, H., et al., 2026, RAA, this issue, xxx

\bibitem[Moretti et al. (2009)]{moretti09} Moretti, A., Pagani, C., Cusumano, G., et al., 2009, \aap, 493, 501

\bibitem[Xie et al. (2024)]{xie24} Xie, W., Cordier, B., Dagoneau, N., et al., 2024, \aap, 683, 60

\end{thebibliography}


\label{lastpage}

\end{document}